\documentclass[a4paper,12pt]{article} 
\usepackage{graphicx} 
\usepackage{graphics} 
\usepackage{amsmath} 
\usepackage{amsfonts}  
\usepackage{amssymb} 

\usepackage[colorlinks=true, pdfstartview=FitV, linkcolor=blue,  citecolor=blue, urlcolor=blue]{hyperref}  

\usepackage{supertabular,booktabs}

\usepackage{amsmath}   

\def\be{\begin{equation}}
\def\ee{\end{equation}}
\def\beq{\begin{equation}}
\def\eeq{\end{equation}}

\begin{document}

%
\title{A catalog of isolated galaxy pairs limited to absolute magnitude -18.5 drawn from HyperLEDA database}
\author{Laurent Nottale$^1$ and Pierre Chamaraux$^2$ \\
{\small $^1$ LUTH, UMR CNRS 8102,  Paris Observatory, 92195 Meudon CEDEX, France}\\
{\small laurent.nottale@obspm.fr}\\
{\small $^2$ GEPI, UMR CNRS 8111, Paris Observatory, 92195 Meudon CEDEX, France}\\ 
 {\small pierre.chamaraux@obspm.fr}}
\maketitle

\begin{abstract}
The present paper is devoted to the construction of a catalog of isolated galaxy pairs extracted from the HyperLEDA extragalactic database. The radial velocities of the galaxies in the pairs are in the range $[3000,16000]$ km.s$^{-1}$. In order to get an unbiased pair catalog as complete as possible, we have limited the absolute magnitude of the galaxies to $M \leq-18.5$).
The criteria used to define the isolated galaxy pairs are the following: 1) Velocity criterion: radial velocity difference between the pair members $\Delta V<500$  km.s$^{-1}$; 2) Interdistance criterion: projected distance between the members $r_p<1$ Mpc; 3) Reciprocity criterion: each member is the closest galaxy to the other one, which excludes multiplets; 4) Isolation criterion: we define a pair as isolated if the ratio $\rho=r_3/r_p$ of the projected distance of the pair to its closest galaxy (this one having a velocity difference lower than 500 km.s$^{-1}$ with respect to the pair) and the members projected interdistance $r_p$ is larger than 2.5. We have searched for these closest galaxies first in HyperLEDA M-limited source catalog, then in the full one.  We have managed not to suppress the small number of pairs having close-by but faint dwarf galaxy companions. The galaxy pair catalog lists the value of $\rho$ for each isolated pair.
This method allows the user of the catalog to select any isolation level (beyond the chosen limit $\rho>2.5$).
Our final catalog contains 13114 galaxy pairs, of which  57\% are fairly isolated with $\rho>5$, and 30 \% are highly isolated with $\rho \geq 10$.
\end{abstract}

\section{Introduction}

      The isolated galaxy pairs represent the simplest gravitational systems of galaxies. Their dynamical study is especially fruitful to estimate the masses and mass-luminosity ratios of their members, and possibly to check the presence of massive haloes and of dark matter (see for instance the basic works by Peterson (1979) \cite{Peterson1979} and by Chengalur et al. (1996) \cite{Chengalur1996}). Such studies need a large catalog of those galaxy pairs, with accurate radial velocities of their members. 
      
     A pioneering catalog of isolated galaxy pairs has been devised by Karachentsev (1972) \cite{Karachentsev1972}; it lists 603 pairs north of the declination $\delta=-3 ^\circ$ and it has been used in several studies. Since then, other catalogs of galaxy pairs have been made available, for instance by Soares et al. (1995) \cite{Soares1995} which completes Karachentsev'one in the Southern hemisphere, and by Karachentsev and Makarov (2008) \cite{Karachentsev2008}, which gathers 509 bound pairs in the Local Supercluster with radial velocities $V_r<3500$ km.s$^{-1}$. 
 More recently the `{\it UGC catalog of isolated galaxy pairs with accurate radial velocities}' by Chamaraux and Nottale \cite{Chamaraux2016} lists 1005 isolated galaxy pairs with a well defined isolation criterion, drawn from Nilson's Uppsala Galaxy Catalog \cite{UGC}.
      
      All these pair catalogs are extracted from old galaxy catalogs with typical size $\approx 10000$ galaxies, issued from the Palomar and ESO Sky Surveys, and therefore they contain $\approx 1000$ pairs or less.
      
       But nowadays, the large galaxy surveys (SDSS, 2MASS, 2dFGRS, etc..) allow to obtain galaxy pair catalogs at a scale of size more than $\approx 10$ times larger. Thus Alonso et al. (2006) \cite{Alonso2006} have used an (unpublished) sample of nearly 13000 pairs extracted from SDSS-DR2 and 2dFGRS in order to study the enhancement of star formation activity in pairs of galaxies due to gravitational interaction. However, as far as we know, there has been no statistical dynamical study performed with such a large sample of pairs.
       
     In the present paper, we construct a catalog of isolated galaxy pairs drawn from the HyperLEDA database \cite{HyperLEDA}, which presently (Sept. 2016) includes  $2714382$ redshifts of galaxies, including as main sources SDSS-DR12 (and previous releases) and 2dF. We have limited our study to radial velocities $2500<V<16500$ km.s$^{-1}$, for which HyperLEDA contains $\approx 250000$ entries, and to pair members of absolute magnitudes $M<-18.5$.
     
     Our main purpose is a dynamical statistical study of the isolated pairs of galaxies, which will be developped in next papers. A specific study will be devoted to the statistical determination of the actual velocity differences and the actual distances between the members of the pairs from the projected values (such a 'deprojection' being needed in order to understand their dynamics).
     
     The paper is organized as follows: section \ref{sec2} is devoted to a short description of the source catalog of galaxies that we have extracted from the HyperLEDA database; in section \ref{sec3}, we present and apply the criteria used in order to define the isolated galaxy pairs; the final catalog of $13114$ galaxy pairs is described in section \ref{sec4}; some of its general statistical properties and corrections for cosmological false pairs are discussed in section \ref{sec5} and we conclude in section \ref{sec6}.

\section{The source catalog extracted from HyperLEDA database}
\label{sec2}

 Presently (2016), the HyperLEDA database contains $4\,964\,207$ redshift measurements for $2\,851\,256$ objects, of which $2\,714\,382$  are galaxies (with a high level of confidence) (\cite{HyperLEDA,Makarov2014} and [Makarov 2016, private communication]). The HyperLEDA redshift database includes the most recent contributions, 10 of which contain more than $\approx 20000$ objects. The main sources are SDSS-DR12 (and previous releases) \cite{SDSSDR12} ($2\,217\,085$ objects) and 2dF ($236\,197$ objects).

 From this database, we have extracted a catalog of galaxies with selected redshifts $2500<V = c z <16500$ km.s$^{-1}$ which contains $249\,685$ entries. Then we have defined a subcatalog limited in absolute magnitude ($M<-18.5$ mag),  which contains $150\,068$ galaxies in the redshift range  $3000<V  <16000$ km.s$^{-1}$ (since our criterion of pair members velocity difference is $\Delta V<500$ km.s$^{-1}$). 
 
 This allows us to construct a pair galaxy catalog whose members are more luminous than $M=-18.5$ and cover the redshift range  $3000<V  <16000$ km.s$^{-1}$. However, the isolation criterion allowing to define the pairs uses the full unlimited galaxy catalog (see herafter).

\section{Definition of isolated galaxy pairs and method}
\label{sec3}

\subsection{Position of the problem}

A galaxy pair is characterized by 6 variables, the three coordinates of the galaxies ``interdistance" $(x, \,y,\, z)$, and the three coordinates of the velocity difference between its members, $(v_x,\, v_y, \, v_z)$. However, only three among these 6 parameters are observationally available. Namely, (assuming the $z$-axis  is oriented from the observer and therefore $(x, \,y)$ is in the plane of the sky), only $x,\,y$ and the radial velocity difference $v_z$ are observable. From the $x$ and $y$ coordinates, we can compute  the interdistance between the pair members projected on the sky plane, $r_p= \sqrt{x^2+y^2}$. In practice, this projected interdistance is derived from the observed angle on the sky $\theta$ between the galaxies and the pair distance $D$, i.e., $r_p=D \, \theta$ (since $\theta\ll 1$ for our pairs, which are such that $V>3000$ km.s$^{-1}$ i.e. $D>43$ Mpc for $h=0.7$.).

Due to these observational constraints, we have to choose our criteria of pair definition from these two parameters, $v_z$ and $r_p$ (in addition to the absolute magnitude limit). We have made the choice to take not too constraining limits (respectively $500$ km.s$^{-1}$ and 1 Mpc), in order not to miss possible real pairs having far apart members. Doing that, we necessarily introduce false (cosmological) pairs, which will be accounted for and excluded in the subsequent analysis.

\subsection{Pair selection}
\label{selection}

We have first extracted from the HyperLEDA database a catalog of galaxies having measured heliocentric reshifts in the range $2500<V <16500$ km.s$^{-1}$. According to our velocity difference criterion ($500$ km.s$^{-1}$), this corresponds to a pair range of redshifts $3000<V <16000$ km.s$^{-1}$.

\begin{figure}[!ht]
\begin{center}
\includegraphics[width=12cm]{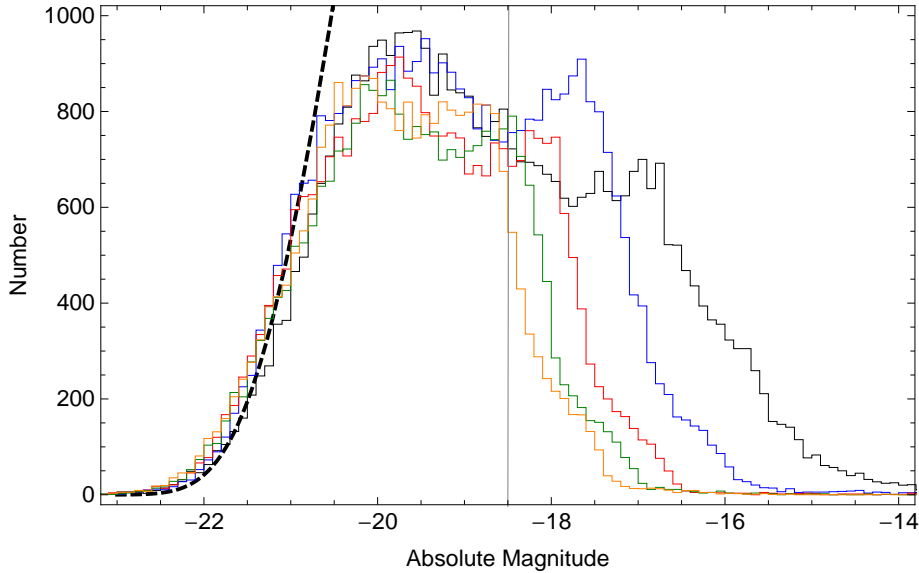}
\caption{{\small  Distance-dependent histogram of absolute magnitudes of galaxies extracted from the HyperLEDA database in the range $3000<V<16000$ km.s$^{-1}$). From right to left, velocity intervals $[3000,\: 7000]$ (black), $[7000,\:10000]$ (blue), $[10000,\:12000]$ (red) , $[12000,\:14000]$ (green) and $[14000,\:16000]$ (orange).  The dashed black line is the Schechter luminosity function \cite{Schechter1976} fitted to the most luminous galaxies.} }
\label{ABSMAG}
\end{center}
\end{figure}

The motivations for this choice are:

 \begin{itemize}

\item $V>3000$ km.s$^{-1}$ allows to use the redshift as distance criterion with enough confidence;

\item $V<16000$ km.s$^{-1}$ is supported by our choice of absolute magnitude limit $M<-18.5$ combined with the observed evolution with distance of the absolute magnitude distribution of galaxies in the HyperLEDA source catalog (see Fig.~\ref{ABSMAG}). Indeed, one can see in this figure that this distribution is complete to an absolute magnitude $M \approx -20.5$ whatever the distance, then becomes incomplete for fainter objects, but in the same way whatever the distance when $M<-18.5$ and $V<16000$ km.s$^{-1}$. For fainter galaxies, the observed distribution becomes highly distance-dependent, and it abruptly falls down at $M=-18.5$ in the farthest velocity bin $[14000-16000]$ km.s$^{-1}$.

\end{itemize}

The data extracted from HyperLEDA include the main object name, alpha and delta coordinates (J2000), total blue magnitude, absolute magnitude, $d25$ diameter, radial velocity, error on radial velocity, distance modulus $mod0$ from redshift independent distance measurement and distance modulus $modz$ from redshift and Hubble law, to which we have added an identification rank (from 1 to $249685$).

Now we list the successive criteria used to construct our catalog of isolated galaxy pairs from the hyperLEDA source catalog defined above. Each successive criterion is applied to the associated sample of pairs and results in a new sample which is more restricted, up to the last criterion which gives the catalog of isolated galaxy pairs.
 
 \begin{itemize}
 
\item (1) First we identify the possible galaxy pair members from the M-limited HyperLEDA source catalog defined above. Such a search insures that radial velocities of those members are in the range $[3000, 16000]$ km.s$^{-1}$ and their absolute magnitudes $M<-18.5$. This allows homogeneity of the sample in the range of distance chosen (see Fig.~\ref{ABSMAG}) and similar luminosity range for the pair members.
 
 \item (2) For each galaxy $A$  in 1), we search its nearest neighbor $B$ included in the M-limited catalog, at a projected distance $r_p<1$ Mpc and with a velocity difference $\delta V_{\rm lim}<500$ km.s$^{-1}$ (see \cite{Chamaraux2016}). That procedure leads to a first set of galaxy pairs.
 
 \item  (3) Reciprocity criterion: a pair defined in 2) is retained only if the closest galaxy to member $B$ is the galaxy $A$ from which the search of the pair has been initiated. Such a procedure allows to exclude the multiplets \cite{Chamaraux2016}. 

 \item (4) Isolation criterion:
  
   We define a galaxy pair as isolated if the ratio $\rho=r_3/r_p$ of the projected distance $r_3$ of the pair to the closest galaxy (this one having a velocity difference lower than 500 km.s$^{-1}$ with respect to the pair) and the members projected interdistance $r_p$ is larger than 2.5. In order to satisfy that criterion, we proceed in two steps:
   
a) We start from the pair sample obtained in 3) and we search for the closest galaxy (denoted \{3\}) as defined above in the M-limited HyperLEDA source catalog; we keep only those pairs for which $\rho>2.5$. 

b) The closest galaxy to each pair kept in a) is then searched in the {\em full} (unlimited) HyperLEDA source catalog, in order to take into account less luminous galaxies (denoted $\{3'\}$). We keep all the pairs with $\rho'>2.5$, and moreover those with $\rho'<2.5$ when the closest galaxy has a luminosity lower than $L_m/10$, $L_m$  being the luminosity of the weakest member of the pair (i.e. $\Delta M>2.5$). Indeed, in that last case, the closest companion has a negligible gravitational influence on the pair at any distance from it.

The galaxy pairs retained by that last criterion constitute our final catalog of 13114 isolated pairs.
  
  \end{itemize}

\section{Final pair catalog: description}
\label{sec4}

The resulting pair catalog (Table \ref{table}) is arranged in order of increasing right ascension of the first member of the pair. It contains, for each pair: column (1) the J2000 $\alpha$ coordinate of the first galaxy of the pair, in decimal hour; (2) the J2000 $\delta$ coordinate of the first galaxy of the pair, in decimal degree; (3) the absolute magnitude of the first galaxy of the pair; (4) the heliocentric radial velocity $V_1$ of the first galaxy of the pair, in km.s$^{-1}$; (5) the uncertainty on $V_1$, in km.s$^{-1}$; (6) the J2000 $\alpha$ coordinate of the second galaxy of the pair, in decimal hour; (7) the J2000 $\delta$ coordinate of the second galaxy of the pair, in decimal degree; (8) the absolute magnitude of the second galaxy of the pair; (9) the heliocentric radial velocity $V_2$ of the second galaxy of the pair, in km.s$^{-1}$; (10) the uncertainty on $V_2$, in km.s$^{-1}$; (11) the absolute value of the radial velocity difference $\Delta V$ between the pair members, in km.s$^{-1}$; (12) the uncertainty on $\Delta V$, in km.s$^{-1}$; (13) the interdistance projected on the sky plane between the two pair members, in Mpc; (14) the ratio $\rho=r_3/r_p$ for the closest galaxy to the pair in the absolute magnitude-limited catalog (see text); (15) the absolute magnitude of this closest galaxy.

A value $\rho=10.$ in column (14) denotes highly isolated pairs with $\rho \geq 10$.

A value $M_3=0.$ in column (15) denotes a missing absolute magnitude for the closest galaxy \{3\} in the HyperLEDA database.

\begin{table}[!ht]
\begin{center}
\caption{The HyperLEDA isolated galaxy pair catalog: examples of pair data. The entries are explained in text. The full catalog will be made available electronically in VizieR.} 
\label{table}
 \fontsize{9}{11}\selectfont
  \setlength{\tabcolsep}{3pt}
\vspace{3 mm}
 \begin{supertabular}{ccccccccccccccccc}
 \hline
  $\alpha_1$ & $\delta_1$ & $M_1$ & $V_1$ & $\varepsilon_{V_1}$ & $\alpha_2$ & $\delta_2$ & $M_2$ & $V_2$ & $\varepsilon_{V_2}$ & $\Delta V$ & $\varepsilon_{\Delta V}$ & $r_p$ & $\rho$ & $M_3$\\ 
   $(1)$ & $(2)$ & $(3)$ & $(4)$ & $(5)$ & $(6)$ & $(7)$ & $(8)$ & $(9)$ & $(10)$ & $(11)$ & $(12)$ & $(13)$ & $(14)$ & $(15)$\\ 
  \hline
0.00124 & -0.08334 & -19.59 & 7125 & 3 & 0.00217 & -0.04059 & -19.51 & 7096 & 3 & 29 & 4 & 0.081 & 10. & 0\\
... & ... & ... & ... & ... & ... & ... & ... & ... & ... & ... & ... & ... & ... & ... \\
0.01549 & -25.9058 & -21.24 & 15083 & 29 & 0.01375 & -25.89882 &
-20.18 & 15317 & 32 & 234 & 43 & 0.094 & 5.46 & -20.\\
... & ... & ... & ... & ... & ... & ... & ... & ... & ... & ... & ... & ... & ... & ... \\
23.99548 & -29.15997 & -20.52 & 8270 & 51 & 23.9847 & -29.12384 & -19.13 & 8715 & 32 & 445 & 60 & 0.302 & 3.05 & -22.2\\
23.99812 & -35.79635 & -18.98 & 15319 & 89 & 23.99964 & -35.77914 & -19.11 & 15140 & 64 & 179 & 110 &
  0.097 & 10. & 0\\
  \hline
\end{supertabular}
\end{center}
\end{table}

\section{Properties of the pair catalog}
\label{sec5}

A detailed analysis of the catalog, in particular regarding statistical deprojections of the velocity difference and distance between members of the pairs, will be done in following papers. Presently we will give some general properties of the catalog.

\subsection{Distance distribution}

\begin{figure}[!ht]
\begin{center}
\includegraphics[width=12cm]{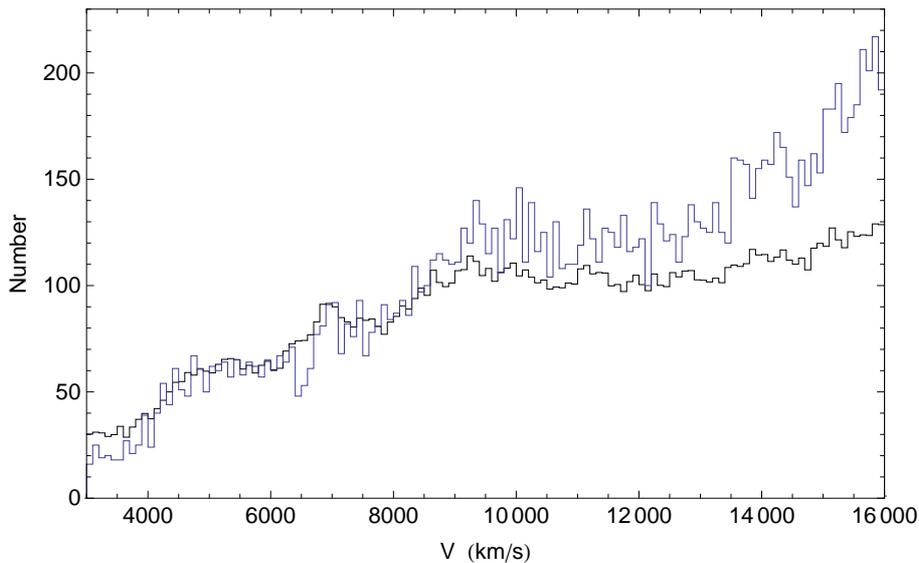}
\caption{{\small  Distribution of the pairs according to cosmological distance, measured by radial velocity (blue line). It is compared to the distribution of all galaxies in the source galaxy catalog (black line) after normalization (by a factor 20) in the range $[3000-10000]$ km.s$^{-1}$. } }
\label{PairDistance}
\end{center}
\end{figure}

The distribution of cosmological radial velocities (and therefore of the distances) of the pairs is shown in Fig.~\ref{PairDistance} and compared to the distance distribution of the galaxies of the source catalog.

The agreement between the two distributions is globally satisfactory, suggesting that the pairs have been correctly drawn at random from the parent HyperLEDA galaxy distribution and that the rate of pairs up to $\approx 10000$ km.s$^{-1}$ $\approx 150$ Mpc is almost constant.

However, beyond $\approx 10000$ km.s$^{-1}$, there is an increasing excess of the observed pair number compared to the source catalog. This effect could be explained by a decreasing efficiency of the isolation criterion, since less and less faint galaxies are taken into account in the research of the closest galaxy to the pair (see Fig.~\ref{ABSMAG}), as well in the complete galaxy catalog as in the absolute magnitude limited one: we have checked that this bias yields the correct order of magnitude for this relative increase.

\subsection{Uncertainties on velocity differences}

\begin{figure}[!ht]
\begin{center}
\begin{tabular}{cc}
 \includegraphics[width=7.5cm]{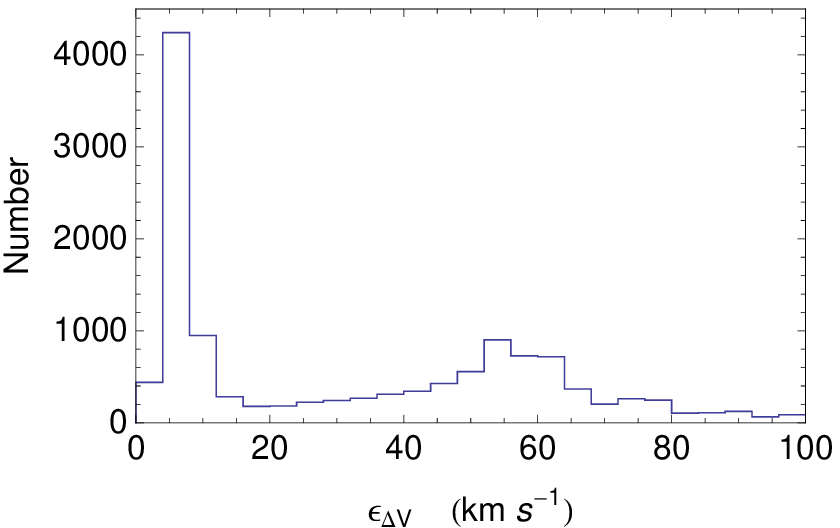} & 
 \includegraphics[width=7.5cm]{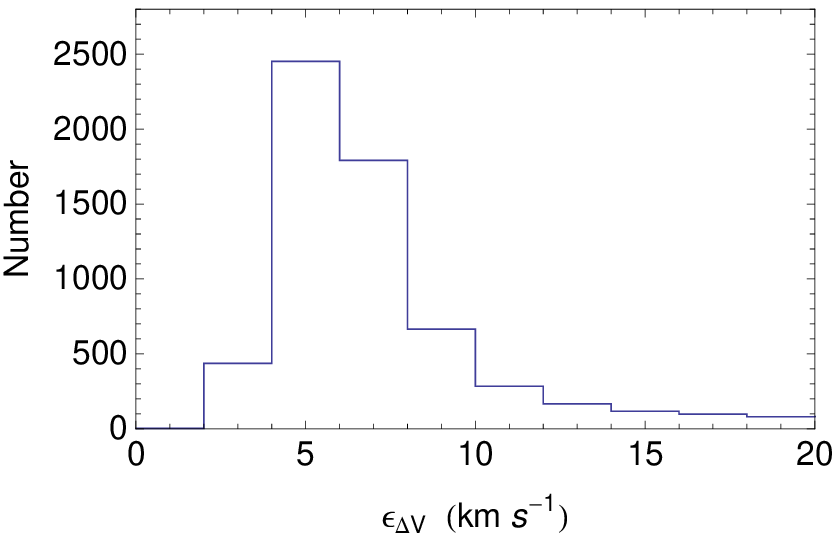} \\
\end{tabular}
\caption{\small{Histograms of the measurement uncertainties $\varepsilon_{\Delta V} =\sqrt{\varepsilon_1^2 + \varepsilon_2^2}$ on the differences between radial velocities of pair members. Left figure: the bin width is 4 km.s$^{-1}$. It shows a sharp peak of accurate velocity measurements at $\approx 5$ km.s$^{-1}$ (radio HI and SDSS measurements) and a second maximum around $\approx 50$  km.s$^{-1}$ due to less accurate optical measurements. Right figure: zoom on the small uncertainty peak. The bin width is 2 km.s$^{-1}$.}}
\label{NCfig3}
\end{center}
\end{figure}

  Figure~\ref{NCfig3} shows the histogram of $\varepsilon_{\Delta V} $. Half of the pairs have very good radial velocity measurements:  6093 pairs have $\varepsilon_{\Delta V} <20$ km.s$^{-1}$ with a sharp peak at 5 km.s$^{-1}$ (see Fig.~\ref{NCfig3} right). For these pairs, the median is $\varepsilon_{\rm med}= 6$ km.s$^{-1}$, which is much lower than the median $\approx 200$ km.s$^{-1}$ of the difference of radial velocities between the members of the pairs.
  
  For the 11461 pairs with $\varepsilon_{\Delta V} <70$ km.s$^{-1}$ (86 \% of the sample) the mean uncertainty is 26 km.s$^{-1}$  and the median as low as 13 km.s$^{-1}$. 
  Therefore we have at our disposal a good quality large sub-catalog which will allow an accurate dynamical analysis.
  
\subsection{Distribution of radial velocity differences between pair members}
\begin{figure}[!ht]
\begin{center}
\begin{tabular}{cc}
 \includegraphics[width=7.5cm]{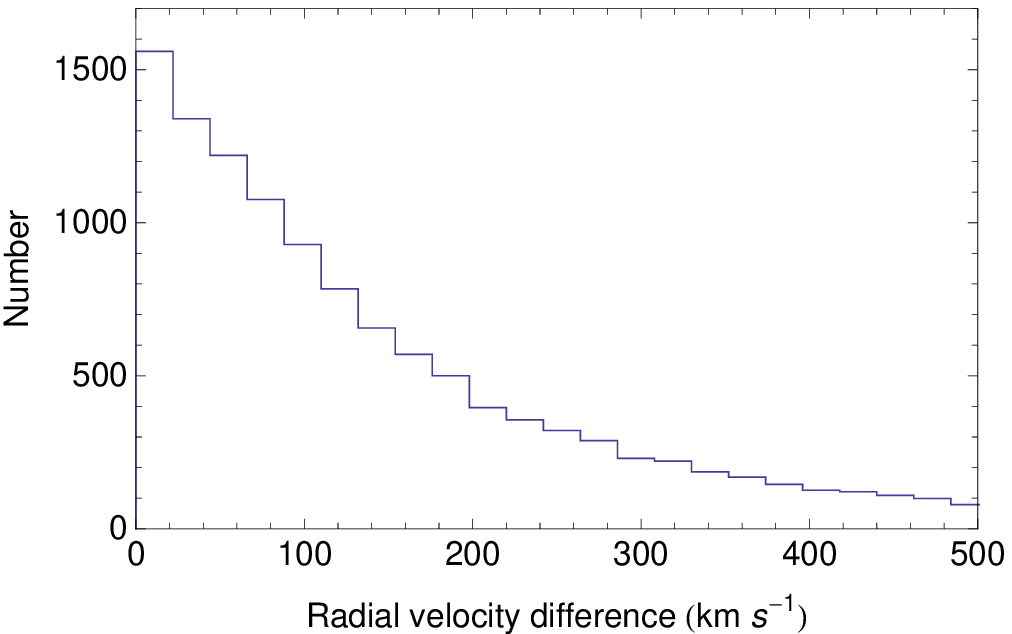} & 
 \includegraphics[width=7.5cm]{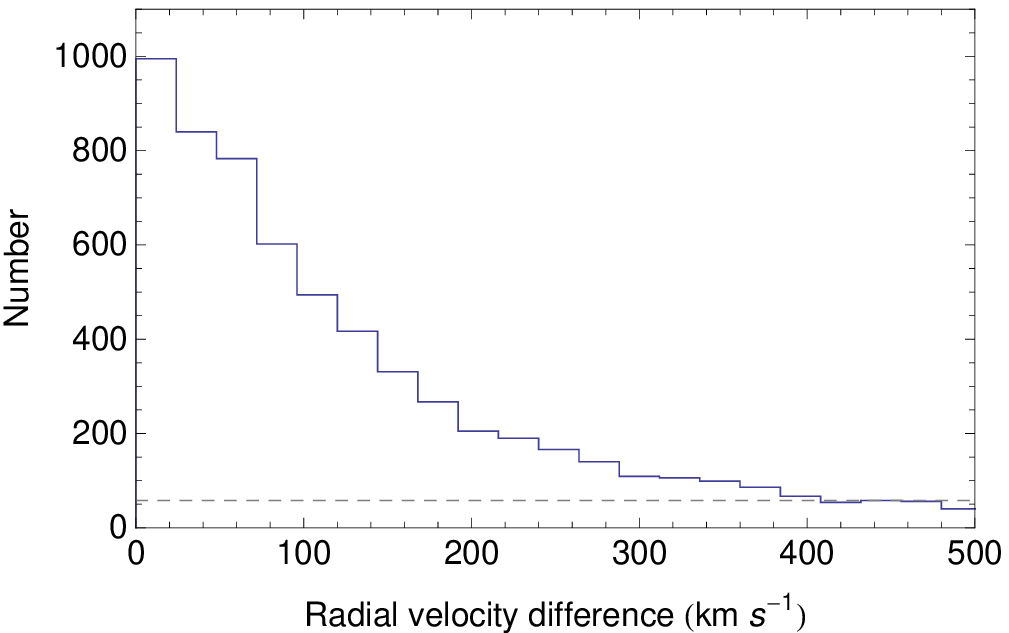} \\
\end{tabular}
\caption{\small{Histograms of the radial velocity differences between pair members. Left figure: uncertainties  $\varepsilon_{\Delta V}<70 $ km.s$^{-1}$, bin 22 km.s$^{-1}$. Right figure: uncertainties  $\varepsilon_{\Delta V}<20$ km.s$^{-1}$, bin 24 km.s$^{-1}$. The histograms are monotonously decreasing, as expected for projected velocities. The dashed horizontal line in the right figure is an estimate of the 'false' cosmological 'pairs' contamination.}}
\label{DV}
\end{center}
\end{figure}

 We give in Fig.~\ref{DV} the histogram of the radial velocity differences between the members of the galaxy pairs, which is monotonously decreasing for increasing radial velocity (within statistical fluctuations). This is theoretically expected, since the probability distribution of the projected velocity for a given $V$ is constant in the interval $[0,V]$. Here, the catalog size is large enough to show a strict monotony despite the fluctuations in $ \approx \sqrt{N}$.

We shall study in more detail this velocity distribution and its `deprojection' in a forthcoming paper.

\subsection{Distribution of interdistances between pair members}

\begin{figure}[!ht]
\begin{center}
\includegraphics[width=12cm]{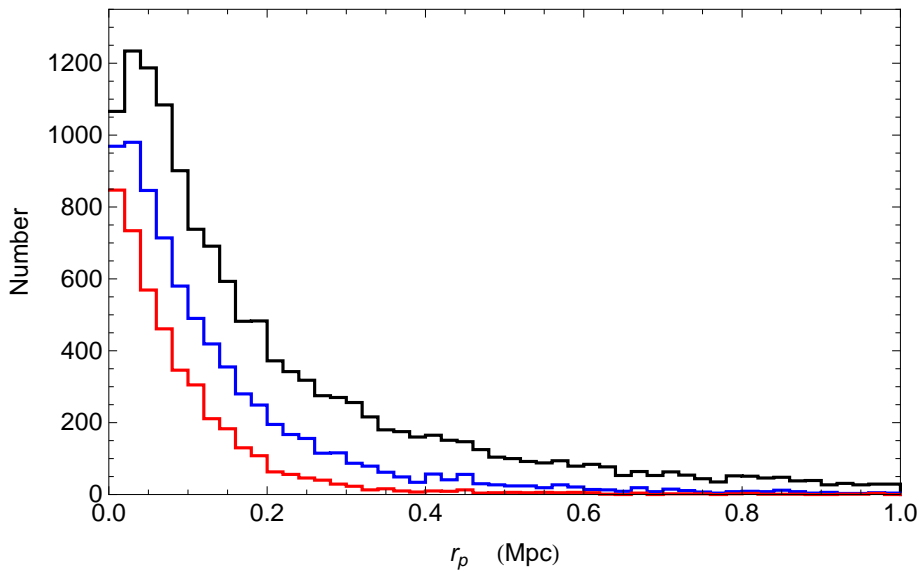}
\caption{{\small Histogram of the projected interdistances between pair members, in bins of 0.02 Mpc. Upper line (black): all data ($\rho>2.5$). Intermediate line (blue): isolated pairs  ($\rho>5$). Lower line (red): highly isolated pairs ($\rho>10$).}}
\label{rp}
\end{center}
\end{figure}

We give in Fig.~\ref{rp} the distribution of projected distances $r_p$ between pair members in our catalog, for three values of the (relative) isolation parameter: $\rho > 2.5$ (weakly isolated pairs), 5 (fairly isolated pairs) and 10  (highly isolated pairs). In all three cases, the distribution is continuously decreasing for increasing values of $r_p$. The distributions approach zero respectively for $r_p \approx$ 0.8, 0.4 and 0.2 Mpc. 

The distribution of $r_p$ in our catalog is well fitted by a $r_p^{-3/2}$ law. This is not in agreement with Turner's result \cite{Turner1976b}, who finds a $r_p^{-1/2}$ law for his sample (defined with differnet criteria), nor with the UGC pair catalog \cite{Chamaraux2016}, for which we find a $r_p^{-1}$ law.

\subsection{Distribution of relative distances to the closest galaxies to the pairs}
\begin{figure}[!ht]
\begin{center}
 \includegraphics[width=10cm]{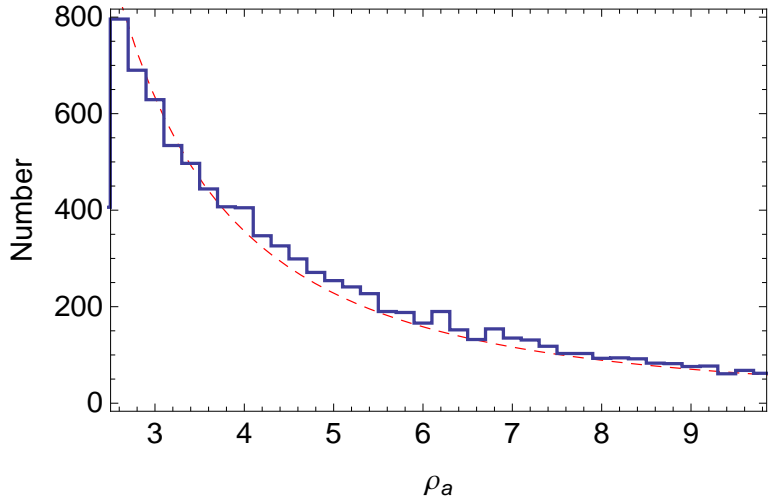} 
\caption{\small{Observed distribution of the relative distances of the nearest galaxies to pair centers (from the absolute magnitude-limited source galaxy catalog) $\rho=r_{a}/r_p$. The proportion of galaxies with $\rho \geq 10.$ is $32$\%. The dashed red line is a $\rho^{-2}$ fit of the distribution.}}
\label{rho}
\end{center}
\end{figure}

We provide in Fig.~\ref{rho} the observed distribution of the relative distances $\rho=r_{3}/r_p$ to the closest galaxies (from the M-limited source galaxy catalog) to pairs. It shows a decreasing behavior as $\rho^{-2}$, in agreement with the distribution found in the UGC pair catalog \cite{Chamaraux2016}.

When we attribute the value $10$ to $\rho$ in our catalog, this means that the nearest galaxy to the pair has not been found in the searched zone, and that it has not been identified (except for some marginal cases); this value in the table means $\rho \geq 10$. The number of  $\rho \geq 10$ is 4268 (32.5\% of the catalog). There are 794 pairs where galaxy $\{3'\}$ is identified ($\rho' \leq 10$) while $\{3\}$ is unidentified. For 2721 pairs, galaxy $\{3'\}$ is closer to the pair than $\{3\}$. Finally a faint proportion of the pairs (350, i.e. 2\%)  have $\rho' \leq 2.5$: these cases correspond to pairs having faint dwarf nearby companions, which have not been excluded despite their mutual proximity since their absolute magnitude difference is $\Delta M >2.5$.

Note that it remains possible that faint galaxies with $M>-18.5$ lie at a distance between $\{3'\}$ and $\{3\}$ when they differ, since we have searched for only the closest galaxy to the pair in both cases. Some pairs should have been excluded from our catalog due to this additional companion $\{4\}$, but since it corresponds to the criteria $\rho_4 \leq 2.5$ and $\Delta M <2.5$ applied to only 350 pairs, their final number is expected to be marginal ($< 2\%$).

\subsection{False (cosmological) ``pairs"}

Some of the pairs are expected to be the result of mere projection effects, their members lying actually at a large relative distance. Indeed, the radial velocity criterion $\Delta V<500$ km.s$^{-1}$ was chosen in order to include very tight pairs, which have large intervelocities due to Kepler's third law. However such a limit for the velocity difference corresponds also to a large cosmological distance,  $d_c=DV/H_0=7$ Mpc (taking a Hubble constant $H_0=70$ km.s$^{-1}$ Mpc$^{-1}$). Then one exceeds our interdistance criterion $r_p<1$ Mpc from $\Delta V=70$ km.s$^{-1}$.

We have attempted to estimate this contamination from cosmological false ``pairs" by using its expected dependence on radial velocity. The volume defined by our criteria is almost a radial cylinder along the line-of-sight, so that the number of false pairs depends linearly on $\Delta V$. Therefore the corresponding rate in the interval $[\Delta V,\Delta V+d\Delta V]$ is constant. 

This is supported by the observed histogram of radial velocity differences (right Fig.~\ref{DV}, uncertainties $\varepsilon_{DV}<20$ km.s$^{-1}$), which shows an almost constant distribution beyond $\Delta V> \approx 400$ km.s$^{-1}$.
The value $400$ km.s$^{-1}$ is a reasonable limit for the maximal velocity difference between members of real pairs. Identifying this flat tail 

(containing $\approx 240$ pairs) to the expected flat cosmological pair distribution yields an upper limit to the false pair contribution.  Since the observed constant rate in the range $\approx 400-500$ km.s$^{-1}$ is $\approx 58$ pairs by bins of 24 km.s$^{-1}$ (see Fig.~\ref{DV}), we estimate by this way the maximal total number of false pairs to be $\approx 1200$, i.e. $<\approx 10$\% of the pair catalog, which is a relatively low contamination.

\section{Conclusion}
\label{sec6}

In this paper, we have carried out the construction of a sample of (weakly to highly) isolated galaxy pairs extracted from the HyperLEDA database. We have limited our catalog to absolute magnitudes $M<-18.5$ and to the redshift range $[3000,16000]$ km.s$^{-1}$.

For this purpose, we have selected accurate quantitative criteria to define isolated galaxy pairs, namely: 1) Low radial velocity difference between the pair members: $\Delta V<500$ km.s$^{-1}$; 2) Small projected distance between the pair members: $r_p<1$ Mpc; 3) Reciprocity, allowing to exclude multiplets; 4) Isolation criterion: we define a pair as isolated if the ratio $\rho=r_3/r_p$ of the projected distance of the pair to its closest galaxy (this one having a velocity difference lower than 500 km.s$^{-1}$ with respect to the pair) and the members projected interdistance $r_p$ is larger than 2.5, first in HyperLEDA M-limited source catalog (closest galaxy $\{3\}$), then in the full one (closest galaxy $\{3'\}$). However we keep the pairs with $\rho>2.5$ and $\rho'<2.5$ when the closest galaxy has a luminosity lower than $L_m/10$, $L_m$  being the luminosity of the weakest member of the pair (i.e. $\Delta M>2.5$), allowing us not to suppress pairs having dwarf galaxy companions. The galaxy pair catalog lists the values of $\rho$ for each isolated pair.

Finally by keeping in the final catalog all the pairs with $\rho$ larger than a rather low limit ($\rho>2.5$), we allow the user of the catalog to choose any isolation criterion beyond this value, defining for example, ``weakly isolated pairs" $2.5< \rho<5$,  ``fairly isolated pairs" ($5< \rho<10$) and ``highly isolated pairs" $\rho \geq 10$. 
Note that Turner \cite{Turner1976} has used the same criterion as that for our ``fairly isolated pairs" ($\rho>5$) to define his pairs.

Our final catalog contains 13114 galaxy pairs, of which 7438 are fairly isolated with $\rho>5$ (57\% of the catalog). Since the source galaxy catalog contains $\approx 150\,000$ objects with $M<-18.5$, this corresponds to a rate of $\approx 10$ \% of galaxies coupled in fairly isolated pairs. We obtained a similar rate for the UGC pairs \cite{Chamaraux2016} with the same isolation criterion $\rho>5$. This rate is also the same as obtained by Karachentsev and Makarov \cite{Karachentsev2008} for isolated pairs (using different criteria). Note also that Gourgoulhon et al \cite{Gourgoulhon1992} have found a proportion of 15\% for galaxies which are members of isolated pairs from a sample of about 4000 nearby galaxies, the isolated pairs being characterized in that study by a well definite overdensity compared to the mean galaxy density.

Finally, 4268 pairs are highly isolated with $\rho \geq 10$ ($32.5 \%$ of the pairs). These pairs include $\approx 5$ \% of the M-limited HyperLEDA galaxies. Once again, this rate is that obtained in the UGC pair catalog, and is the same as in the pioneering 1972 Karachentsev's catalog \cite{Karachentsev1972}.

    The distribution of the pairs versus their cosmological distance is compatible with that of the HyperLEDA source galaxies, with a small bias at large distances. The proportion of ``false" cosmological pairs, i.e. galaxies which seem close to each other due to projection effects, is estimated to be low (less than 10 \% of the catalog).
      
      Thus we have at disposal a quite fair and large sample, which allows significant statistical studies. This pair catalog construction, following the extraction of the UGC pair catalog (containing 1005 pairs), is a second step which has allowed to gain a factor $>10$ in the number of pairs, using the now available large databases.
      
      In a forthcoming paper, we intend to perform a dynamical study of those galaxy pairs, in order to obtain the masses of their members, their mass-luminosity ratios, and possibly to check the presence of massive haloes and of dark matter in them. Such a work needs a statistical derivation of the actual velocity differences between the members and of their actual interdistances from their known measured projections, a task which will be developped and used in the future analysis of our galaxy pairs catalogs.\\

{\bf Acknowledgements.}
We acknowledge the use of the HyperLEDA database (http://leda.univ-lyon1.fr) and we gratefully thank Dr. Dmitry Makarov for kindly providing us with up to date informations about the database.\\


\end{document}